\documentclass[onecolumn,amssymb, sort&compress, nobibnotes, aps, prb]{revtex4}
\setcitestyle{numbers,square}
\usepackage[bookmarks=false]{hyperref}
\usepackage[utf8]{inputenc}
\usepackage[vietnamese=nohyphenation]{hyphsubst}
\usepackage[vietnamese,english]{babel}
\usepackage[T5]{fontenc}
\usepackage{array}
\usepackage{amssymb}
\usepackage{amsmath}
\usepackage{graphicx}
\usepackage{color}
\usepackage{physics}
\usepackage{float}
\usepackage{textcomp}
\usepackage{gensymb}
\usepackage{hyperref}
\usepackage{multirow}
\usepackage{tabularray}
\usepackage[normalem]{ulem}
\usepackage{comment}
\usepackage{longtable}

\hypersetup{
     colorlinks   = true,
     citecolor    = blue
}
\renewcommand{\vec}[1]{\boldsymbol{#1}}
\begin{document}
\noindent

\title{Superconductivity and correlated phases in non-twisted bilayer and trilayer graphene} 

\author{Pierre A.\ Pantale\'on}
\email{ppantaleon@uabc.edu.mx}
\affiliation{$Imdea\ Nanoscience,\ Faraday\ 9,\ 28049\ Madrid,\ Spain$}
\author{Alejandro Jimeno-Pozo}
\affiliation{$Imdea\ Nanoscience,\ Faraday\ 9,\ 28049\ Madrid,\ Spain$}
\author{H\'ector Sainz-Cruz}
\affiliation{$Imdea\ Nanoscience,\ Faraday\ 9,\ 28049\ Madrid,\ Spain$}
\author{\foreignlanguage{vietnamese}{Võ Tiến Phong}}
\affiliation{Department of Physics and Astronomy, University of Pennsylvania, Philadelphia PA 19104, USA}
\author{Tommaso Cea}
\affiliation{$Department\ of\ Physical\ and\ Chemical\ Sciences,\ University\ of\ L'Aquila,\ via\ Vetoio,\ Coppito,\ 67100\ L'Aquila,\ Italy$}
\author{Francisco Guinea}
\email{paco.guinea@gmail.com}
\affiliation{$Imdea\ Nanoscience,\ Faraday\ 9,\ 28049\ Madrid,\ Spain$}
\affiliation{$Donostia\ International\  Physics\ Center,\ Paseo\ Manuel\ de\ Lardizabal\ 4,\ 20018\ San\ Sebastian,\ Spain$}

\date{\today}

\begin{abstract}
The discovery of a very rich phase diagram in twisted bilayer graphene~\cite{Cetal18,Cetal18b} renewed the interest into the properties of other systems based on graphene. An unexpected finding has been the observation of superconductivity in non-twisted graphene bilayers and trilayers~\cite{Zhou2022SCBG,Zhou2021SuperRTG,zhang2022spin}. In this perspective, we give an overview of the search for uncommon phases in non-twisted graphene systems. We first describe results related to the topic before the aforementioned experiments~\cite{Zhou2022SCBG,Zhou2021SuperRTG,zhang2022spin} were published. Then, we address the new experimental findings which have triggered the recent interest in the problem. Lastly, we analyze the already numerous theory works studying the underlying physical processes~\cite{G21}.
\end{abstract}

\maketitle

\section{Unusual phases in non-twisted graphene stacks. Early results.}
Graphite is considered a semimetal, with a highly complex Fermi surface, and a low density of states at the Fermi level. To a first approximation, graphite can be seen as a stack of two dimensional graphene layers. The low energy electronic properties of each graphene layer are well described by the two dimensional Dirac equation~\cite{M69,NGPNG09}. This model shows that the density of states at the Fermi level vanishes, suggesting that electron interactions do not play a major role.

The isolation, and posterior detailed study~\cite{Netal04,Netal04b}, of individual graphene layers lead to a search for effects of electronic interactions in single layer graphene, and also in bilayer graphene and in graphene multilayers. One of the most striking effects of the electronic interactions in single layer graphene is the renormalization of the Fermi velocity which characterizes the Dirac cones~\cite{GGV94,Eetal11}, similarly to the renormalization processes in Quantum Electrodynamics. The degeneracy between Landau levels of monolayer graphene is also broken by interactions~\cite{Zetal06,G11}.

The generalization of the Dirac equation to graphene (Bernal) bilayers~\cite{MF06,KM13} replaces the two touching Dirac points by two touching parabolic bands. This structure leads to a logarithmically divergent susceptibility~\cite{NNPG06}. In the renormalization group language, this result implies that the screened Coulomb interaction is a marginally relevant perturbation, which leads to a variety of instabilities~\cite{MBPM08,V10,NL10,LAF12}, including superconductivity. Superconductivity can also arise at high electron dopings, when the Fermi energy is close to a van Hove instability~\cite{NLC12,Metal10}. Other graphene multilayers show interesting electronic bands, see, for instance~\cite{K10}. In particular, an infinite rhombohedral stack is a Dirac semimetal~\cite{AMV18}, which implies that flat two-dimensional bands exist at the top and bottom surfaces, which can lead to correlated states~\cite{PMC17}. Localized states at surfaces are likely in finite rhombohedral stacks~\cite{ZSMM10,SMKF19}. Finally, infinite graphene stacks can show stacking disorder (defected graphite), and, at the position of the defects, quasi-two-dimensional behavior may emerge~\cite{GNP06,AG08,ZJFNM11,MCM21,GSF22}. Superconductivity in rhombohedral graphite has been predicted theoretically~\cite{KHV11,KIHH13,V18}.

On the experimental side, superconductivity was observed in the so called graphite intercalation compounds, made up of metallic and graphene layers~\cite{DD02,Setal15,Takada2016}, with critical temperatures up to $\sim 10-12$K. Superconductivity at higher temperatures, $\sim 30$K has been found in fullerenes, $C_{60}$, crystals doped with alkaline ions~\cite{Rosseinsky1991,Kelty1991,Chakravarty1991,Hebard1991,H92}. In both cases, the amount of charge transferred to the carbon systems is considerable, $\sim 0.1 - 1$ $e$/atom, much larger than typical electrostatic dopings achieved in few layer graphene stacks. The electronic bands of graphite intercalation compounds and alkali doped fullerenes are quite different from the bands of graphene and graphite, and there is no clear evidence for unconventional superconductivity, although the effect of electron-electron interactions has been invoked in the doped fullerenes~\cite{CFCT02,NSCA15,Calandra2005SC,Wang1991SC,Cantaluppi2018,Haddon1986C60,Erwin1991K3C60}.

In addition, some early results suggested the existence of correlated phases in Bernal bilayers~\cite{Wetal10,FTWS12,Betal12,Varlet2014LiquidBG,Letal14,Moriyama2019SCbil} and also in $ABC$ trilayers~\cite{Chen2019Tril,Chen2019GateTunableMottRTG, Chen2020TunableCorrelatedChernRTG}. Some observations of electronic gaps in transport measurements can be due to domain walls in the bilayers~\cite{SGGNG14}. Unusual phases have also been reported in other graphene trilayers~\cite{Letal14b,Zibrov2018GullyTG,Chen2019Tril}, and also in thicker multilayers~\cite{NKKMM16,NKSM18,Ketal21}. 

Finally, it is worth remarking that a number of works report phenomena compatible with superconductivity in bulk graphite~\cite{E13,Eetal14,Eetal18,E19,Scheike2012,Layek2022,Ariskina2022,Retal22}. It is unclear the role that two dimensional electron gases trapped at stacking defects can play in these observations~\cite{GNP06,AG08,ZJFNM11,E19,Ariskina2022,MCM21,GSF22}.\\

\section{Superconductivity in Bernal bilayer and rhombohedral trilayer graphene}
\subsection{Bilayer graphene}

\begin{figure}[h]
    \includegraphics[width=17.5cm]{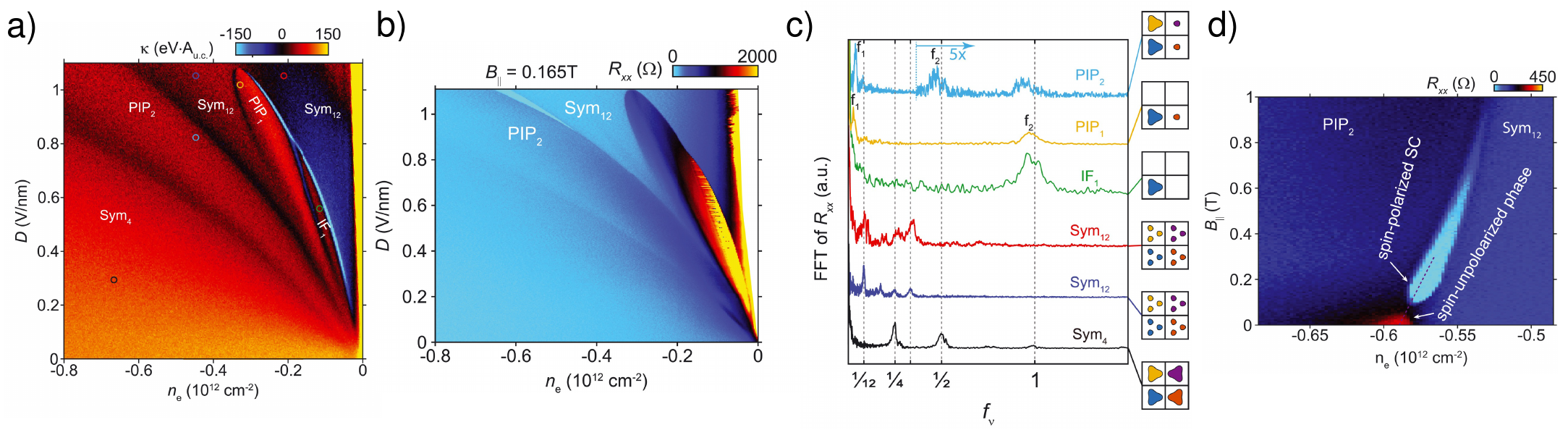}
    \caption{Superconductivity in bilayer graphene. (a) Inverse compressibility without magnetic field, as a function of electron doping $n_e$ and displacement field $D$.\ (b) Longitudinal resistivity ($R_{xx}$) at $T=10$ mK and $B_\parallel=165$ mT, showing a superconducting sleeve in bright cyan.\ (c) Fast Fourier transform of the magneto-resistance versus frequency normalized to $n_e$, measured at circled points in (a).\ The insets shows the inferred Fermi surfaces for each phase. (d) $R_{xx}$ versus $B_\parallel$ and $n_e$ at $D=1.02$ V/nm.\ Figures extracted from Ref.~\cite{Zhou2022SCBG}.}    
    \label{fig:FigZhouetal22}
\end{figure}

Superconductivity in Bernal bilayer graphene was discovered by Zhou $et.\ al.$~\cite{Zhou2022SCBG}. This material is the most common allotrope of graphene because its stacking is the structural ground state of two graphene layers~\cite{Yan2011Bernal,Lee2010Bernal}. The experiments in Ref.~\cite{Zhou2022SCBG} consist of inverse compressibility and magneto-resistance measurements on bilayer graphene encapsulated in non-aligned hBN and graphite gates. Upon applying a displacement field perpendicular to the layers, a gap opens at charge neutrality, while the bands develop a `Mexican hat’ dispersion, leading to van Hove singularities that promote strong correlations. In this setup, Ref.~\cite{Zhou2022SCBG} shows a map of the inverse compressibility $\kappa=\partial\mu/\partial n_e$, where $\mu$ is the chemical potential,  as a function of hole doping $n_e$ and displacement field $D$.\ The map reveals a series of dips in $\kappa$, which delineate different regions in the phase space defined by $n_e$ and $D$, see Fig.~\ref{fig:FigZhouetal22}(a). Furthermore, measuring the magneto-resistance yields the frequency of quantum oscillations normalized to $n_e$, $f_\nu$, which is the fraction of the Fermi surface enclosed by a cyclotron orbit. The results show that $f_\nu$, and hence the number of pockets in the Fermi surface, is different in each region. Together, these phenomena indicate a cascade of transitions between flavour-symmetry-breaking metallic phases, which resemble those observe in twisted graphene stacks~\cite{zondiner2020cascade,wong2020cascade,choi2021correlation} and RTG~\cite{Zhou2021HalfMetRTG}, as well as Lifshitz transitions which change the topology of the Fermi surface. Subsequent experiments have studied in great detail the correlated phases of bilayer graphene~\cite{barrera2022cascade,seiler2022quantum}. In particular, on top of the flavour polarized metals, Ref.~\cite{seiler2022quantum} reports the observation of several new phases, such as correlated metals with linear-in-temperature resistivity, a Wigner crystal, and a novel Wigner-Hall crystal, an electron crystal which has a quantized Hall conductance.

Upon applying an in-plane magnetic field, a superconducting sleeve appears near the boundary between two of these phases, as shown in Fig.~\ref{fig:FigZhouetal22}(b), for dopings near where a van Hove singularity is expected. One of these phases, dubbed `Sym$_{12}$', appears to have twelve Fermi pockets (four flavours with three pockets each due to trigonal warping effects, see Fig.~\ref{fig:FigBands}(f)) and the other, `PIP$_2$', a symmetry-broken phase with two large and two small pockets, one per flavour, which may be nematic. Superconductivity manifests as vanishing resistivity, and a Fraunhofer pattern in the presence of a magnetic field, conclusive evidence of phase coherence. The critical temperature, deduced from non-linear transport and Berezinskii-Kosterlitz-Thouless theory, is $T_{c}\approx26$ mK. At this temperature, a magnetic field of $40$ mT would break the pairs of a conventional spin-singlet superconductor, since the Zeeman effect would overcome the superconducting gap. Superconductivity in bilayer graphene survives at fields over ten times larger than this Pauli limit, of $B\gtrsim 500$ mT, which is strong evidence of spin-triplet pairing. 

The emergence of superconductivity is better understood by comparing transport experiments at zero and finite magnetic field, throughout the region of phase-space that becomes superconducting. At zero magnetic field, the sleeve between PIP$_2$ and Sym$_{12}$ is a state distinct from both, with higher resistance, and, most likely, insulating, see Fig.~\ref{fig:FigZhouetal22}(d). Its resistance has non-linear behaviour and drops abruptly at a threshold in the applied current. Turning on and increasing a magnetic field (in- or out-of-plane) reduces this threshold until it vanishes, giving rise to superconductivity. This suggest that the resistive state is spin-unpolarized and that the superconducting state emerges from it when the magnetic field polarizes the spin.

A relevant measure of the level of disorder in the material is the ratio between the superconducting coherence length, estimated to be $\xi\approx250$ nm and the electronic mean free path, of $l_M\approx5$ $\mu$m, comparable to the dimensions of the device. The ratio is $d={\xi}/{l_M}<0.05$, which implies an ultra-clean regime of superconductivity, in which pair-breaking effects due to non-magnetic impurities are weak, and which is therefore compatible with the unconventional spin-triplet pairing inferred from the rest of the data. A similar ratio is estimated in the experiments described next. 

\begin{figure}[h]
   \includegraphics[width=15cm]{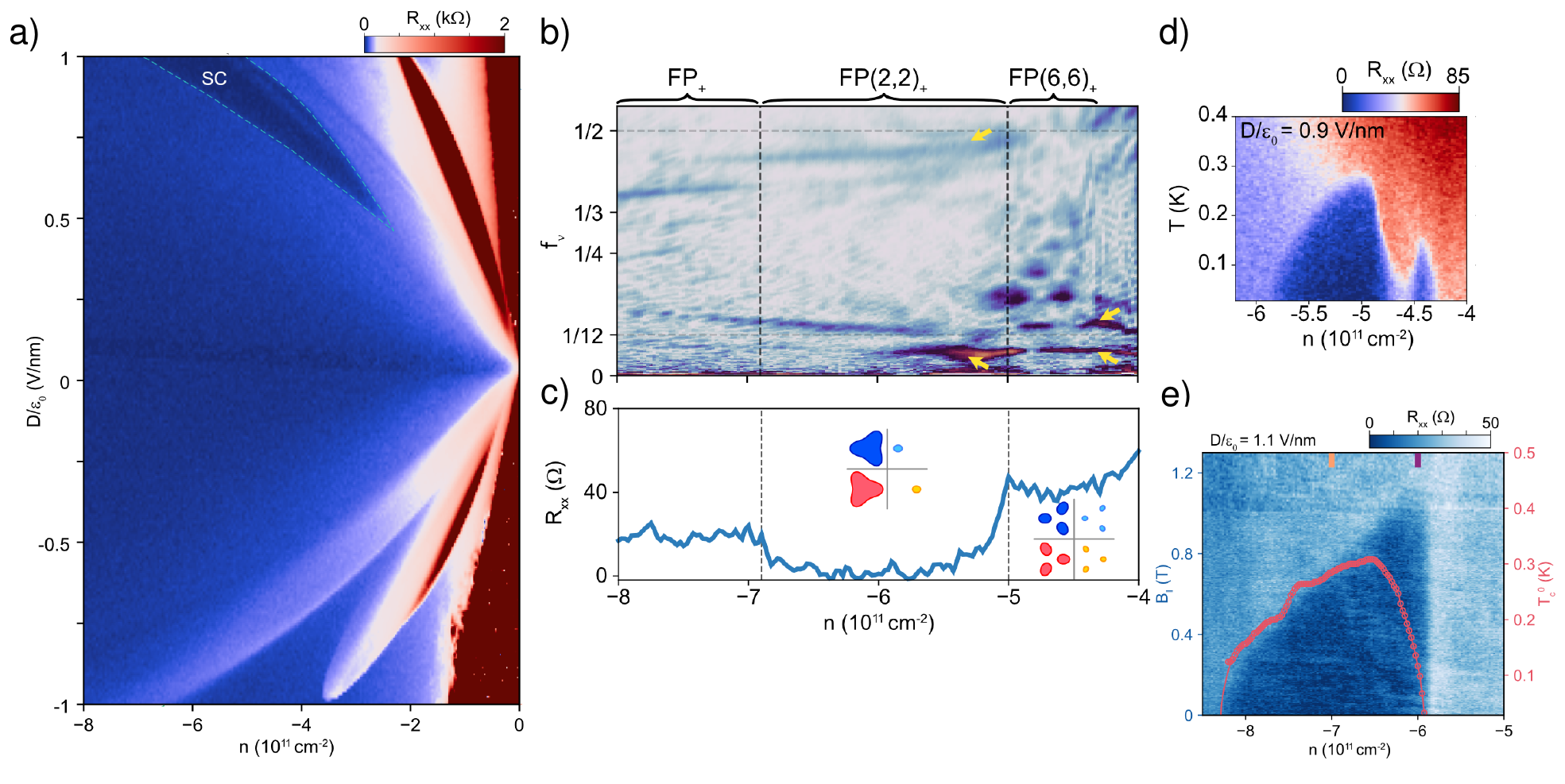}
    \caption{Superconductivity in bilayer with WSe$_2$. (a) $R_{xx}$ versus hole doping and displacement field, at zero magnetic field, showing a superconducting sleeve in dark blue. (b) Fast Fourier transform of the magneto-resistance versus $n_e$. (c) Cut of the resistivity at $D=1$ V/nm, insets show the Fermi surfaces inferred from (b). (d) $R_{xx}$ as a function of doping and temperature for $D=0.9$ V/nm. (e) $R_{xx}$ versus doping and in plane magnetic field, at $D=1.1$ V/nm. Red circles indicate the zero-field $T_c$. Figures extracted from Ref.~\cite{zhang2022spin}.}   
    \label{fig:FigZhangetal22}
\end{figure}

The second set of ground-breaking experiments on bilayer graphene have been done by Zhang $et.\ al.$~\cite{zhang2022spin}. The main idea was to stabilize superconductivity by placing tungsten diselenide (WSe$_2$) on top of the bilayer, a strategy that has already proved fruitful in twisted bilayer graphene, yielding superconductivity at record low twist angles~\cite{arora2020superconductivity}. For bilayer graphene, Ref.~\cite{zhang2022spin} reveals that the interplay with WSe$_2$ leads to a significant, global enhancement of superconductivity: it appears even in the absence of a magnetic field, its critical temperature increases by an order of magnitude, to $T_c\approx$ 260 mK and the filling range of the superconducting sleeve is similarly extended, compare Fig.~\ref{fig:FigZhangetal22}(a) to  Fig.~\ref{fig:FigZhouetal22}(a). There is compelling evidence that these effects are due to Ising spin-orbit coupling (SOC) induced by WSe$_2$. Note that in the absence of WSe$_2$ the SOC in bilayer graphene is negligible~\cite{Konschuh2012SOCBil}.

A key aspect of Fig.~\ref{fig:FigZhangetal22}(a) is the asymmetry with respect to the $D$ field direction. In particular, superconductivity appears just at positive $D$. Theory predicts that Ising SOC only couples to the top layer, which is next to WSe$_2$~\cite{gmitra2017proximity,khoo2017demand}. For hole doping, a positive $D$ field pushes the wavefunction towards WSe$_2$, letting the carriers benefit from Ising SOC. For similar values of $n_e$, the phase at negative $D$ fields is a resistive state that disappears upon applying magnetic fields, which is most likely the same phase from which superconductivity emerges in Ref.~\cite{Zhou2022SCBG}. It appears that this is also the phase that here, at positive $D$ field, competes with superconductivity, cutting through the superconducting region and splitting it in two, see Fig.~\ref{fig:FigZhangetal22}(d). A resistive state, flanked by superconducting domes, is reminiscent of the situation observed in e.g. twisted bilayer graphene~\cite{Cetal18b}. These observations support the idea that Ising SOC contributes to  superconductivity in the system.

Quantum oscillations confirm that the phase diagram observed in~\cite{zhang2022spin} at negative $D$ field is almost identical to the one observed in bilayer graphene in~\cite{Zhou2022SCBG}, which shows a transition between the PIP$_2$ and Sym$_{12}$ phases.\ In contrast, for positive $D$ field, the data suggests that, due to band splitting induced by Ising SOC, the phase adjacent to PIP$_2$ has six large and six small Fermi pockets, so it is similar to PIP$_2$ but in the trigonal warping regime, see Fig.~\ref{fig:FigZhangetal22}(b) and (c).\ Another difference is that superconductivity emerges from PIP$_2$ after Ising SOC presumably selects a symmetry-broken state that enables pairing, while in bilayer graphene encapsulated in hBN the parent state seems to be another one with higher resistance, see Fig.~\ref{fig:FigZhouetal22}(d).

The critical magnetic field differs starkly on both sides of the superconducting dome, as shown in Fig.~\ref{fig:FigZhangetal22}(e).\ The low $n_e$ side violates the Pauli limit by a factor of six, while the high $n_e$ barely does, perhaps hinting at a nuanced evolution of the Fermi surface with doping within the dome.\ A possibility is that the system develops Ising superconductivity, which promotes Cooper pairs of type $\ket{K^{+},\uparrow; K^{-},\downarrow}$, locking the spin and valley degrees of freedom and making the state very resilient to in-plane magnetic fields.\ However, as pointed in Ref.~\cite{zhang2022spin}, the moderate critical fields observed may be more consistent with an Ising SOC-induced inter-valley coherent phase.

\subsection{Trilayer graphene}

\begin{figure}[h]   
    \includegraphics[width=17.5cm]{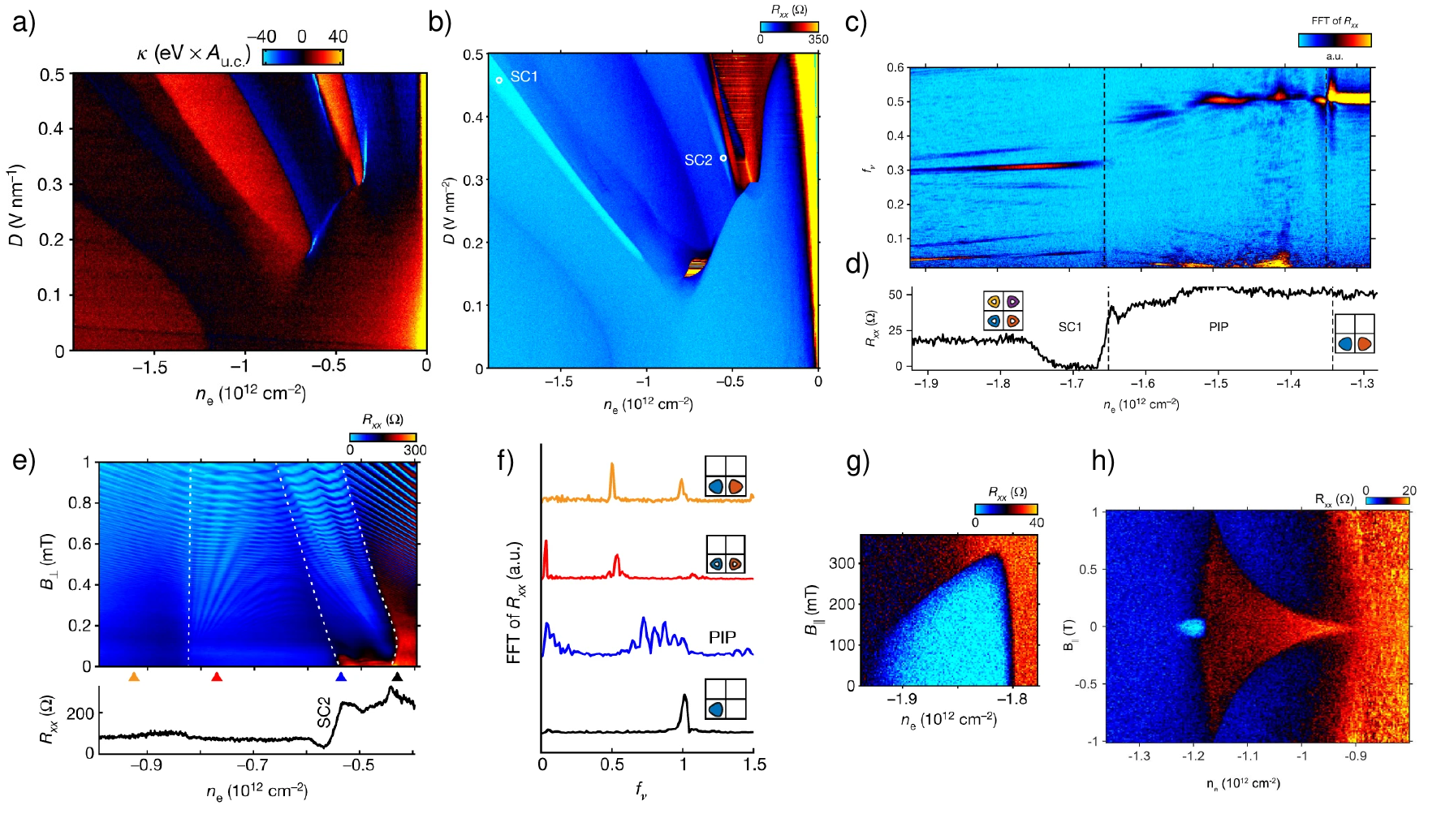}
    \caption{Superconductivity in RTG. (a) Inverse compressibility as a function of $n_e$ and $D$. (b) $R_{xx}$ versus $n_e$ and $D$, showing two superconducting sleeves SC1 and SC2 in bright cyan. (c) Fast Fourier transform of the magneto-resistance as a function of $n_e$ near SC1 and frequency $f_{\nu}$. (d) $R_{xx}$ versus $n_e$ near SC1, at zero magnetic field. (e) Magneto-resistance versus $n_e$ near SC2 (top) and $R_{xx}$ versus $n_e$ near SC2, at zero magnetic field (bottom).
    (f) Fast Fourier transform of (e), along the coloured arrows. (g) $R_{xx}$ as a function of $n_e$ and temperature, for SC1. (h) $R_{xx}$ versus $n_e$ and in-plane magnetic field, at the partially flavour polarized phase adjacent to SC1.
    Figures extracted from Ref.~\cite{Zhou2021HalfMetRTG} and Ref.~\cite{Zhou2021SuperRTG}.}    
    \label{fig:FigZhouetal21}
\end{figure}

The experiments by Zhou {\it et. al}.~\cite{Zhou2021HalfMetRTG,Zhou2021SuperRTG} in rhombohedral trilayer graphene (RTG) reported unambiguous observations of correlated behaviour~\cite{Zhou2021HalfMetRTG} and superconductivity~\cite{Zhou2021SuperRTG}  in a crystalline graphene system. Rhombohedral $ABC$ stacking is a metastable three layer allotrope of graphene, with Bernal $ABA$ being the ground state, although recent results suggest $ABC$ may be stable~\cite{GPGWA21}. RTG is similar to BBG in that a perpendicular displacement field opens a gap and flattens the low-energy bands, setting the stage for strong correlations, although it is worth noting that RTG has flat bands even in the absence of a $D$ field. In this material, inverse compressibility and magneto-resistance measurements also uncover a cascade of flavour-symmetry breaking transitions, see Fig.~\ref{fig:FigZhouetal21}(a), including a spin-polarized half-metal phase and a spin- and valley-polarized quarter-metal. Ref.~\cite{Zhou2021HalfMetRTG} uses a Stoner model, similar to the one in Ref. \cite{zondiner2020cascade}, to explain the cascade at positive filling. At negative fillings, shown in Fig.~\ref{fig:FigZhouetal21}(a), the situation is more nuanced, due to the interplay between flavour symmetry breaking and Lifshitz transitions, which gives rise to several partially flavour polarized phases. Superconductivity has been found in two regions, SC1 and SC2, of the phase space defined by $n_e$ and $D$, as shown in Fig.~\ref{fig:FigZhouetal21}(b). On top of vanishing resistivity, non-linear transport in the presence of a magnetic field yields Fraunhofer patterns, definitive evidence of superconductivity. The estimated critical temperatures are $T_{c1}=106$ mK for SC1 and $T_{c2}=50$ mK for SC2. Like in BBG, the superconducting sleeves in RTG appear close to flavour-symmetry-breaking transitions. 

Quantum oscillations data near the SC1 superconducting sleeve, shown in Fig.~\ref{fig:FigZhouetal21}(c), reveals that it arises from a symmetric state, Sym$_{4}$, in which each flavour has an annular Fermi surface, and which is adjacent to a phase with a complex Fermi surface, spin-unpolarized and partially valley-polarized, probably with large and small Fermi pockets, like PIP$_2$ in BBG. On the other hand, SC2 emerges at the boundary between a spin-polarized, valley-unpolarized half-metal and a partially-flavour polarized state, which also appears to have large and small Fermi pockets, see Fig.~\ref{fig:FigZhouetal21}(f). Further differentiating these two superconducting regions, SC1 shows only a small Pauli limit violation, and its $T_c$ dependence on $B$ is compatible with conventional spin-singlet pairing. Still, it is worth noting that the critical magnetic field is very asymmetric with respect to $n_e$, like in BBG-$\text{WSe}_2$, compare Fig.~\ref{fig:FigZhouetal21}(g) and Fig.~\ref{fig:FigZhangetal22}(e). In stark contrast, SC2 violates the Pauli limit by an order of magnitude. Moreover, spin-singlet pairing for SC2 is precluded by the fact that its parent state is spin-polarized, so the two spin orientations are separated by the exchange interaction energy ($\sim 1-10$ meV), over two orders of magnitude greater than the superconducting gap. Therefore, there is strong evidence that SC2 is a spin-triplet superconductor. Just like in BBG, an estimate of the level of disorder gives $d\sim0.05$, so the system is clean enough to support unconventional spin-triplet pairings.

\begin{figure}[h]
     \includegraphics[scale=1.0]{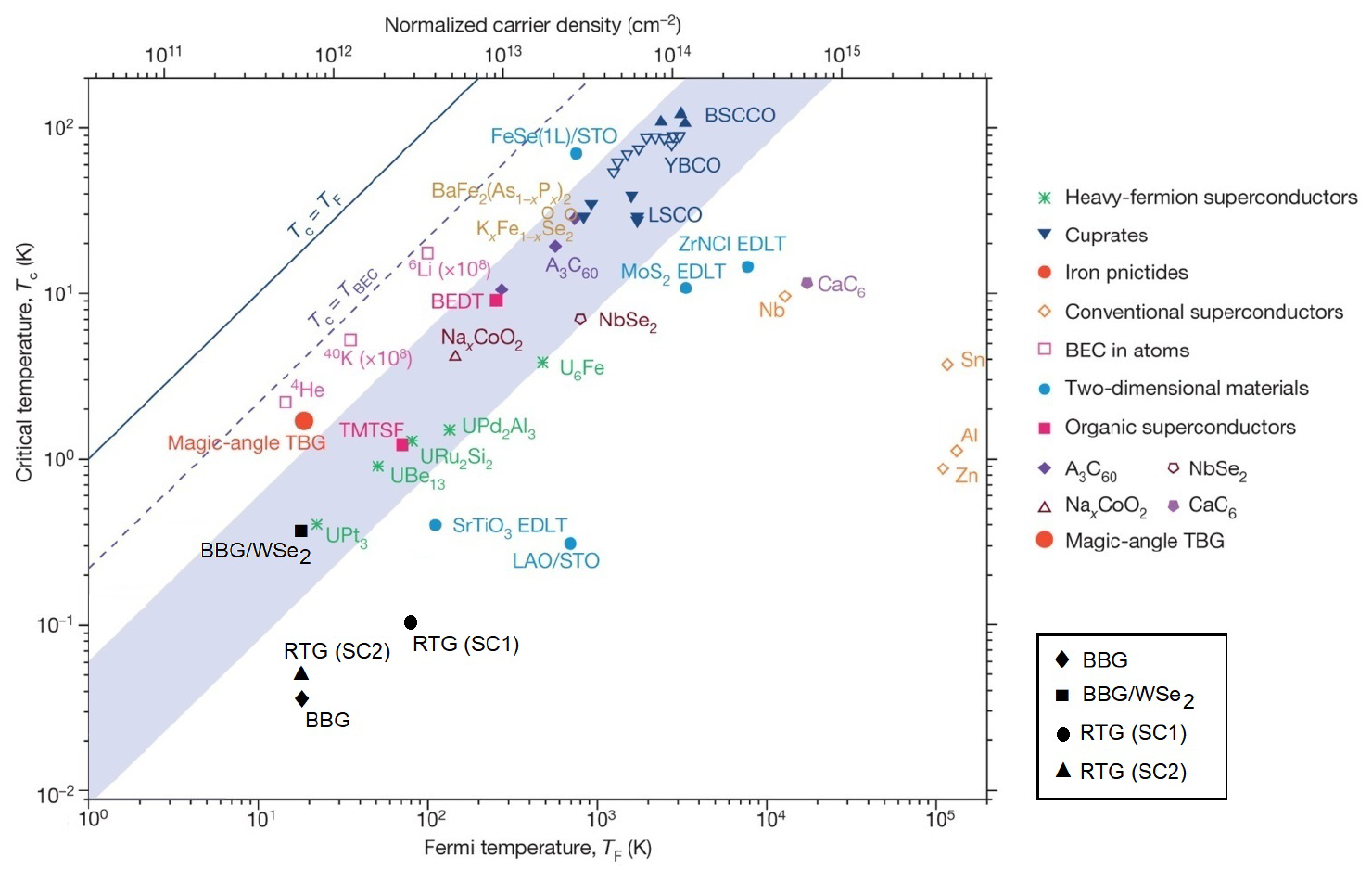} 
    \caption{Logarithmic plot of critical temperature $T_c$ versus Fermi temperature $T_F$ for various superconductors. Black shapes correspond to non twisted graphene systems. Adapted from Ref.~\cite{Cetal18b}   
    }
    \label{fig:FigCrit} 
\end{figure}

\subsection{Comparison to twisted stacks}
It is illustrative to compare the superconductivity and correlated phases observed so far in twisted and non-twisted stacks. Some of the similar features shared by both types of stacks are that (i) superconductivity occurs at very low carrier densities. The density in twisted bilayer graphene is determined by the number of electrons which can be accommodated into the central bands, while in non twisted systems the density is fixed by the states available in the flat regions of the low energy bands. In both cases these estimates give  $n_{e} \sim 10^{12} \text{ cm}^{-2}$.
 Knowledge of $n_{e}$ allows for the classification of interactions according to their strength, and reveals that the long range Coulomb interaction, $V_C\approx e^2 k_F\propto e^2 \sqrt n_e$, is larger than the bandwidth and it is parametrically larger than other interactions \cite{CPWG22}, which scale as $n_e$. Also, superconductivity: (ii) violates the Pauli limit, indicating that the pairing is, at least in part, spin-triplet; (iii) is adjacent to correlated insulators~\cite{Cetal18b,park2021tunable,hao2021electric,Cao2021Pauli,zhang2021ascendance,park2022robust,Zhou2022SCBG,zhang2022spin} and (iv) is modified when the graphene stack is in proximity of a WSe$_2$ monolayer~\cite{arora2020superconductivity,zhang2022spin}. In addition, there are (v) cascades of flavour-symmetry breaking transitions~\cite{zondiner2020cascade,wong2020cascade,Zhou2021HalfMetRTG,Zhou2021SuperRTG,barrera2022cascade,seiler2022quantum,Zhou2022SCBG,zhang2022spin}; (vi) Quantum Anomalous Hall phases~\cite{serlin2020intrinsic,Zhou2021HalfMetRTG,seiler2022quantum} and (vii) strange metals with linear-in-temperature resistivity~\cite{polshyn2019large,seiler2022quantum}.

The most notable differences include: (i) $T_c\sim1$ K in twisted stacks, and $T_c\sim10-100$ mK in non-twisted stacks. Since the densities are similar, this means that twisted stacks are in a regime of stronger coupling, see Fig.~\ref{fig:FigCrit}. (ii) Superconductivity in non-twisted stacks appears only when applying a displacement field. (iii) The relevant bands are isolated from the rest of the spectrum only in twisted stacks with an even number of layers. (iv) Angle disorder, strain and lattice relaxation have a serious impact on twisted stacks. (v) In proximity of a WSe$_2$ monolayer, $T_c$ is enhanced in BBG, while in twisted bilayer graphene superconductivity is stabilized at a lower angle, but $T_c$ does not increase. (vi) Some phenomena have only been observed in twisted stacks, e.g. Chern insulators~\cite{chen2020tunable,nuckolls2020strongly,saito2021hofstader,wu2021chern,pierce2021unconventional,stepanov2021competing,xie2021fractional}, nematicity~\cite{cao2021nematicity}, Pomeranchuk effect~\cite{rozen2021entropic,saito2021isospin} or re-entrant superconductivity with a magnetic field~\cite{Cao2021Pauli}, while others like the Wigner and Wigner-Hall crystals have only been seen in BBG~\cite{seiler2022quantum}. 

\section{Theoretical Models}

\subsection{Non-interacting band structures}
\begin{figure}[h]
     \includegraphics[scale=0.36]{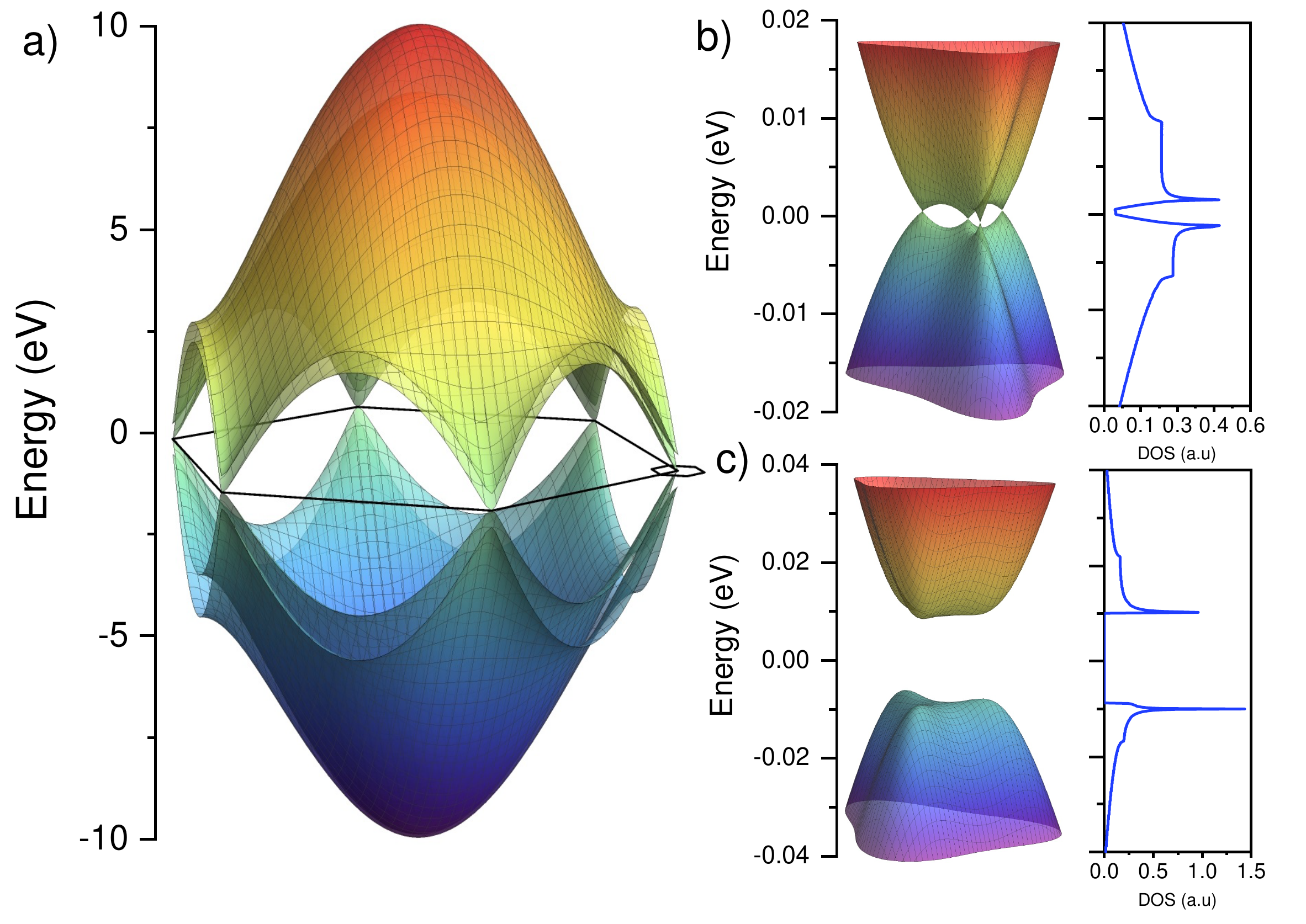} \includegraphics[scale=0.36]{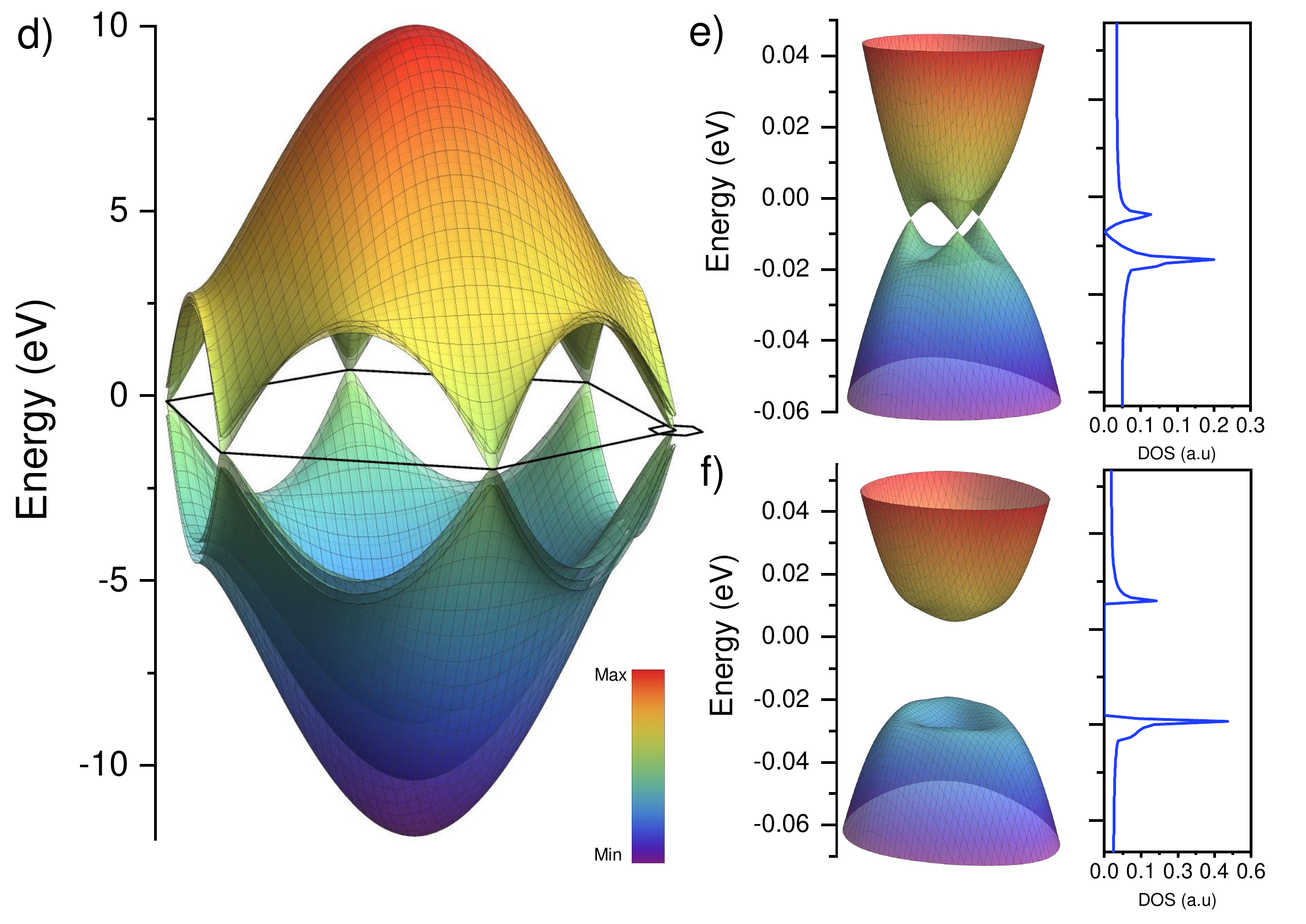} 
    \caption{Band structure in the full Brillouin zone for: a) Bernal bilayer graphene and d) rhombohedral trilayer graphene. The large hexagon near zero energy is the corresponding Brillouin zone. The small hexagon is the Dirac point where the low energies of the corresponding system are calculated for b) and e) zero and c) and f) $20$ meV perpendicular displacement field. At the side of the low energy bands is the corresponding density of states. The model and parameters for the figures are given in Ref.~\cite{KM13} 
    }
    \label{fig:FigBands}
\end{figure}
Both RTG and BBG become superconductors at very low electronic densities $n_{e} \sim 10^{12} \text{ cm}^{-2}$  when an electric field is applied perpendicular to the graphene system. From the theoretical point of view is universally accepted that the band flattening resulting from the displacement field and the consequent van Hove singularities enhance the role of interactions. The study of the electronic properties is usually performed with low energy continuum models with the exception of Ref.~\cite{CPPG22} which studied the superconductivity using a tight binding model. The non-interacting tight binding band structure~\cite{KM13}, low energy regions and density of states of RTG and BBG is shown in Fig.~\ref{fig:FigBands}. In BBG there are two vertically aligned graphene layers so that atoms belonging to the sublattice $A$ of layer $1$ lie over the atoms in the sublattice $B$ of layer 2. Similarly to monolayer graphene, BBG is a semi-metal in which the low energy bands touch at the Dirac points, but with parabolic instead of linear dispersion~\cite{Netal06,MF06}. This band touching makes the charge susceptibility and other susceptibilities diverge~\cite{NNPG06}, which leads to symmetry-broken phases~\cite{MBPM08,V10,NL10,LAF12,zhang2012bil,cvetkovic2012crit,martin2010local}.  As a perpendicular electric field is applied, inversion symmetry is broken, and the Dirac points connecting the bands become gapped. As a result, the local band dispersion can be nearly quenched. As shown in Fig.~\ref{fig:FigBands}(b) and (c), the low energy bilayer bands acquire a `Mexican hat' profile~\cite{zhang2009gap}.

Rhombohedral trilayer graphene is a semi-metal with an approximate cubic band degeneracy at the zone corners~\cite{M69,DD02,AG08,ZSMM10,SMKF19,K10,Betal11,KHV11,KIHH13,Letal14,PMC17} where optimal  parameters have been reported in the literature~\cite{K10,ZSMM10,Zibrov2018GullyTG,Shi2020RTG}. Close to the Dirac points, the cubic degeneracy resulting from diagonalizing the Hamiltonian splits into three Dirac cones, creating a trigonally-warped Fermi surface, shown in Fig.~\ref{fig:FigBands}(e). A perpendicular electric field breaks inversion symmetry, and these Dirac points acquire a finite mass, Fig.~\ref{fig:FigBands}(f). As a result, the local band dispersion is flattened, generating a van Hove singularity that favors the emergence of correlated electronic phases.

\begin{center}
\begin{longtblr}[
    caption = {Overview of theories of superconductivity in bilayer and trilayer graphene systems, according to their proposed pairing mechanism, estimated critical temperature $T_c$ and symmetry of the superconducting order parameter (OP).},
    label = {tab:mergedTable}
]{|c|p{5cm}|p{2.5cm}|p{4cm}|p{2.5cm}|}
\hline
System               & Pairing mechanism                                                      & $T_{c}$ (mK)                       & OP symmetry                                                                        & OP sign change \\ \hline\hline
\SetCell[r=5]{} BBG                 & Phonons \cite{CWSS22,CWSS22b}                         & 1000 - 2000    (no $e$-$e$ interactions), 20-500 (with $e$-$e$ interactions)                   & Spin-triplet~\cite{pol}, valley singlet ($f-$wave), intra-valley: singly degenerate, $A$.                     & Most likely, no                         \\ 
                     & Short range, momentum independent interactions, proximity to broken symmetry phases \cite{SR22}& ---&Spin-triplet ($f-$wave).  & ---\\
                     & Quantum critical modes \cite{DCL22}                   & 35                          & Spin-triplet, valley-singlet $s-$wave or spin-triplet, valley-triplet ($f-$wave).               & ---                         \\ 
                     & Screened long-range Coulomb \cite{JSCPG22}            & 10                          & Spin-triplet, valley-singlet ($f-$ wave), intra-valley: singly degenerate, $A$, or doubly degenerate, $E$. & Yes                         \\ 
                     & Short-range Coulomb \cite{C22}                        & 20                          & Spin-triplet, valley-singlet ($f-$wave).                                                         & Most likely, no                         \\ \hline \hline
\SetCell[r=2]{} BBG-$\text{WSe}_{2}$ & Phonons + substrate tunneling \cite{chou2022enhanced} & ---                             & Mixture of spin-singlet, valley-triplet ($s-$wave) and spin-triplet, valley-singlet ($f-$wave)                              & ---                         \\ 
                     & Screened long-range Coulomb \cite{JSCPG22}            & 12-40 & Ising superconductor, intra-valley: singly, $A$, or doubly degenerate, $E$.                     & Yes                         \\ \hline \hline

\SetCell[r=9]{} RTG                 & Phonons \cite{Chou2021AcousticRTG}  & 500 - 1600  & Spin-singlet, valley-triplet (unpolarized normal state), spin-triplet, valley-singlet (polarized normal state)  & Most likely, no \\ 
& Short range, momentum independent interactions, proximity to broken symmetry phases \cite{Szabo2022RTG}& ---&Spin-triplet ($f-$wave).  & ---\\
                     & Flavour fluctuations \cite{ZL21}  & ---  & Spin-singlet, valley-triplet, singly degenerate, $A$ & Most likely, no \\
                        & Flavour fluctuations \cite{Chatterjee2022}  & ---  & Dependent on details of interactions & Dependent on details of interactions \\
                     & Intervalley fluctuations \cite{ZV21} & ---                         & Spin-singlet, valley-triplet & ---
                     \\
                      & Screened long-range Coulomb \cite{Cea2022SCKLRTG,JSCPG22}            & 65-140                        & Spin-triplet, valley-singlet ($f-$ wave), intra-valley: singly degenerate, $A$, or doubly degenerate, $E$. & Yes                         \\ 
                      & Screened long-range Coulomb \cite{Qin2022FRG_RTG} & 0 - 2000, depending on dielectric constant                          & Two pockets in each valley, spin singlet, $d-$ wave (doubly degenerate), or spin triplet, $p-$wave (doubly degenerate), depending on inter-pocket couplings & Most likely, yes\\ 
                      & Screened long-range Coulomb \cite{GHSB21} & Depends on coupling constant, $\lambda$. $T_{c}$ covers a broad range.  & Depends on details of intervalley coupling. & Yes\\
                      & Few momentum independent interactions \cite{Lu2022RG_RTG} & ---                         & Spin triplet, valley-singlet or spin-singlet, valley-triplet, depending on sign of inter-valley Hund coupling & ---\\            
                      & Short-range Coulomb \cite{ Dai2022QMonteCarlo}  & ---  & Spin-singlet, $d-$wave (at half filling only)  & --- \\
                      & Short-range Coulomb \cite{C22}                        & 20-140                          & Spin-triplet, valley-singlet ($f-$wave).                                                         & Most likely, no \\
                      & Short-range Coulomb \cite{DHZLM21}    & --- & d + $\mathrm{i}$d. & ---
                      \\
                     \hline \hline
RTG-$\text{WSe}_{2}$  & Screened long-range Coulomb \cite{JSCPG22}            & 65-70   & Ising superconductor, intra-valley: singly degenerate,$A$, or doubly degenerate, $E$.  & Yes \\
                     \hline
\end{longtblr}
\end{center}

\subsection{Superconducting mechanisms}
\begin{figure}[h]
    \includegraphics[scale=0.6]{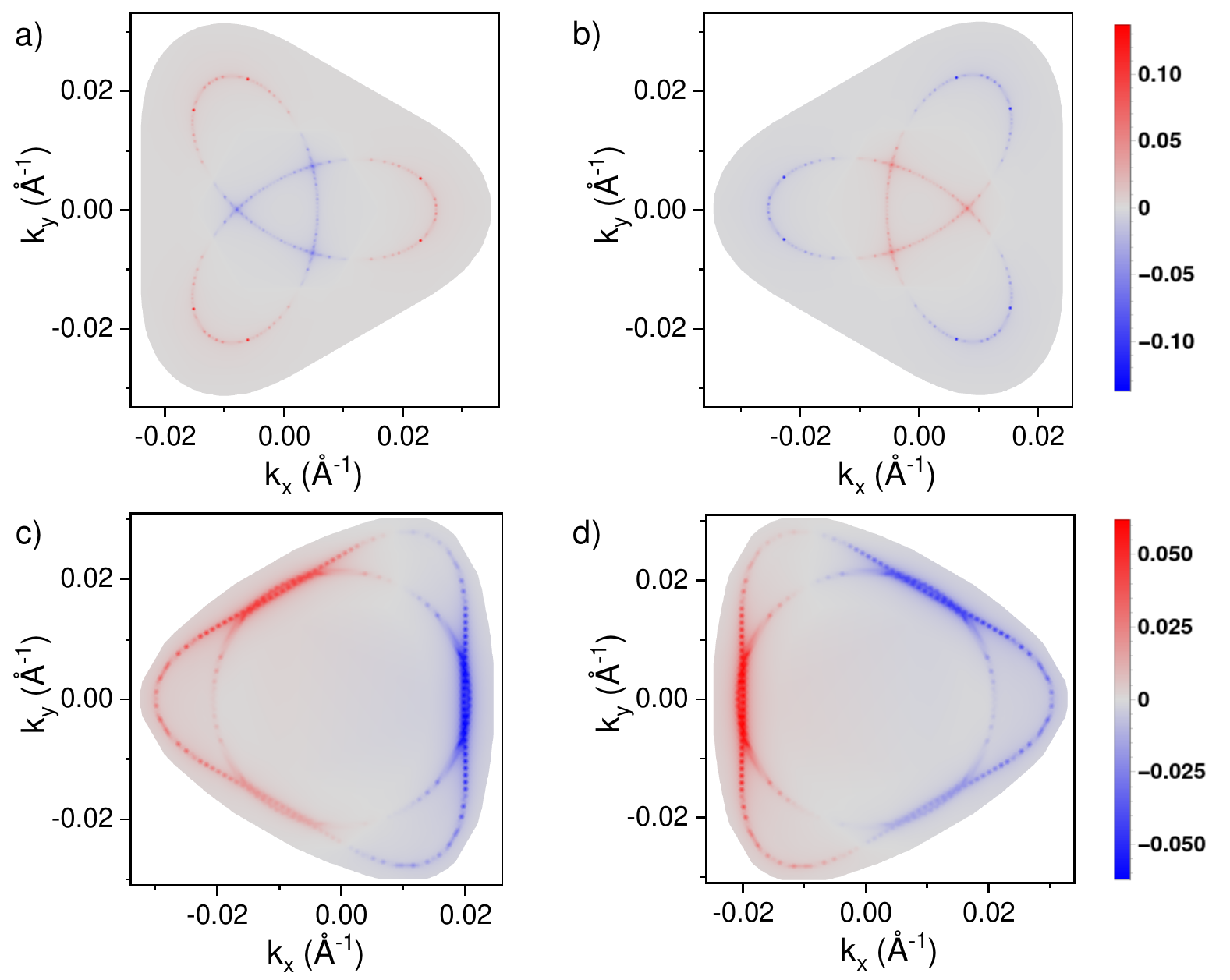}
    \caption{(a) Superconducting order parameter (OP) of BBG in valley $K^{+}$, (b) in valley $K^{-}$, near the hole-doped van Hove singularity, with $T_{c}\approx10$ mK.\ Within a single valley, the OP peaks along the edges of the Fermi surface and changes sign between the inner and outer edges.\ (c) OP of RTG in valley $K^{+}$, (d) in valley $K^{-}$, near the hole-doped van Hove singularity and with $T_{c}\approx65$ mK, showing intensity stripes along the edges of the Fermi surface.\ $C_3$-symmetry is broken because the intravalley symmetry is $E$, degenerate.\ In both materials, the OP changes sign between valleys, which means that BBG and RTG are valley-singlet, spin-triplet superconductors. Data replotted from Ref.~\cite{JSCPG22}}
    \label{fig:KL_OP}
\end{figure}

It is worth first to describe some general features of the superconducting state. The symmetries of the hamiltonian constrain the possible superconducting order parameters. The existence of two valleys implies that the order parameter can be even or odd with respect to valley exchange. If we neglect spin-orbit coupling, the Pauli principle implies that a valley symmetric order parameter is a spin-singlet, while a valley antisymmetric order parameter is a spin-triplet. The lowest symmetry that the bilayer and trilayer in a perpendicular field has around the $K$ and $K'$ points is a ${\cal C}_3$ rotation. This symmetry allows for two representations, a non degenerate one, $A$, and a doubly degenerate one, $E$. Spin-orbit coupling mixes valley and spin exchange. The simplest case where the spin-orbit coupling only affects the out of plane spin direction, $s_z$, leads to Ising superconductivity, where the spin and valley degrees of freedom are locked~\cite{ising}. Finally, depending on the sign and relative strength of intra- and inter-valley interactions, the order parameter can show changes of sign and nodes at the Fermi surfaces, an example is shown in Fig.~\ref{fig:KL_OP}. 

Using this analysis of the superconducting state, a classification of different theoretical models is given in Table~\ref{tab:mergedTable} for both Bernal bilayer and $ABC$ trilayer. Three general pairing mechanisms have been considered: i) electron-phonon coupling~\cite{CWSS22,CWSS22b,chou2022enhanced,Chou2021AcousticRTG}, ii) electronic fluctuations associated to the proximity of broken symmetry phases~\cite{SR22,DCL22,Szabo2022RTG,ZL21,Chatterjee2022,ZV21}, and iii) direct repulsive electron-electron interactions, either the long range Coulomb potential~\cite{JSCPG22,Cea2022SCKLRTG,Qin2022FRG_RTG,Lu2022RG_RTG,GHSB21} or short range interactions~\cite{C22,Dai2022QMonteCarlo,DHZLM21}. These pairing mechanisms have been studied using a variety of formalisms: i) standard BCS calculations~\cite{CWSS22,CWSS22b,chou2022enhanced,Chou2021AcousticRTG}, ii) diagrammatic methods inspired by the Kohn-Luttinger \cite{KL65} formalism~\cite{ZV21,Cea2022SCKLRTG,JSCPG22,C22}, the renormalization group~\cite{Qin2022FRG_RTG,Lu2022RG_RTG}, phenomenological electronic susceptibilities describing nearby broken symmetry phases~\cite{SR22,DCL22,ZL21,ZV21}, or numerical Monte Carlo analyses~\cite{Dai2022QMonteCarlo}. Different approximations require the knowledge of different number of parameters.

The changes of sign of the order parameter at the Fermi surfaces determine the robustness of the superconducting phase against elastic scattering. When the order parameters has different signs in the two valleys, short range scattering is pair breaking. If the order parameter changes sign within the pockets in each valley, long range scattering is also pair breaking. It is worth mentioning that long range repulsive interactions can lead to a superconducting state because the screened interaction is less repulsive at $\vec{q} = 0$ than at finite values of $\vec{q}$, so that an order parameter with sign changes is allowed~\cite{MC13}.

Finally, we compare the results of the theoretical proposals collected in Table \ref{tab:mergedTable} with the experimental findings in Refs.~\cite{Zhou2022SCBG, zhang2022spin,Zhou2021SuperRTG}. In BBG, which has a measured critical temperature of $T_c\approx26$ mK, the theories in which pairing is due to purely electronic interactions \cite{SR22, DCL22, JSCPG22, C22} yield critical temperatures and superconducting dome widths of the same order of magnitude as the experiments, while in the theory where pairing is mediated by phonons \cite{CWSS22,CWSS22b} the predicted critical temperatures and the width of the superconducting dome are larger than in experiments. Identical conclusions apply to RTG. When a monolayer of WSe$_{2}$ is placed on top of BBG, the critical temperature goes up by an order of magnitude, reaching $T_c\approx260$ mK. In this setup, Ref. \cite{chou2022enhanced} shows that virtual tunnelling to the substrate can favour the attraction between two holes and boost the phonon-driven superconducting $T_c$, although an estimate of the enhancement is not given. Ref. \cite{JSCPG22} proposes a Kohn-Luttinger-like mechanism for superconductivity and shows that, due to reduced screening of the Coulomb interaction at intermediate momenta, Ising spin-orbit coupling enhances the superconducing $T_c$ by a factor of 4 in BBG. The associated order parameter shows locking between the spin and valley degrees of freedom.

Most theories either predict, or are compatible with spin-triplet pairing, in agreement with the Pauli limit violations seen in experiments. Furthermore, some proposals based on the screened long-range Coulomb interaction \cite{Cea2022SCKLRTG,JSCPG22} predict the form of the superconducting order parameter, which shows intravalley nodes and sign changes, see Fig.~\ref{fig:KL_OP}. As mentioned above, these sign changes make superconductivity fragile against long-range disorder, and may explain why it has eluded discovery until now.

\subsection{Other theoretical results.}
The analyses described in the previous subsection give also information on the nature of the normal state, especially those ones which assume that superconductivity emerges from a broken symmetry phase. In addition, the role of the electron-electron interaction in the normal state of RTG has bee studied in~\cite{CCB22}, where the effects of strong local interactions are described using Dynamical Mean Field Theory. Non superconducting broken symmetry phases in RTG are analyzed phenomenologically in~\cite{HWQWBM22}. The information that the optical conductivity gives on correlated phases in RTG has been studied in~\cite{JMS22}. Superconductivity has also been predicted to exist in rhombohedral tetralayer graphene~\cite{CWSS22b, Ghazaryan2022Multilayers}.

\section{Conclusions}
The crystalline nature of bilayer and trilayer graphene is a crucial advantage with respect to twisted systems, which suffer from twist angle disorder~\cite{uri2020} and strains~\cite{kazmierczak2021strain}. These issues place strict demands on the fabrication process and hinder experimental reproducibility~\cite{lau2022reproducibility}. In contrast, bilayer and trilayer graphene are mechanically stable, easy to fabricate, and have very low disorder and strain. Given the complexity of the phase diagrams to elucidate, bypassing the issue of disorder may prove decisive in the pursuit of a unified theory of superconductivity and strongly correlated phases in graphene based superconductors. 

The observation of superconductivity and other non trivial phases in non-twisted graphene bilayers and trilayers suggests that other graphene based compounds, suitably modified, can also support correlated electronic phases. Non-twisted stacks, like the ones discussed here, probably can be tuned to show flat bands. Note that, on the other hand, not all twisted graphene stacks show superconductivity~\cite{Cetal20,Letal20TDBG}. 

Superconductivity has not been found yet in twisted transition metal dichalcogenides (TMD's), or in TMD heterobilayers, although there is ample evidence of other correlated phases~\cite{MS22}. 

Finally, it is worth considering whether the mechanism which triggers superconductivity is the same in twisted and non-twisted carbon compounds.

\section{Acknowledgments.}
We are thankful to Andrey\ V.\ Chubukov, Jos\'e A. Silva-Guillen, Elena Bascones and Gerardo Naumis for illuminating discussions.\ We acknowledge support from the Severo Ochoa programme for centres of excellence in R\&D (Grant No.\ SEV-2016-0686, Ministerio de Ciencia e Innovaci\'on, Spain); from the European Commission, within the Graphene Flagship, Core 3, grant number 881603 and from grants NMAT2D (Comunidad de Madrid, Spain), SprQuMat (Ministerio de Ciencia e Innovaci\'on, Spain) and (MAD2D-CM)-MRR MATERIALES AVANZADOS-IMDEA-NC. V.T.P. acknowledges support from  the Department of Energy under grant DE-FG02-84ER45118, the NSF Graduate Research Fellowships Program, and the P.D. Soros Fellowship for New Americans.


\begin{thebibliography}{146}
\bibitem[pol()]{pol}
The superconductivity in this model is assumed to emerge from a spin polarized
  normal state.
\bibitem[isi()]{ising}
The valley and spin degrees of freedom can also be locked in the normal phase,
  when interactions lead to broken symmetry metallic phases\cite{LPS22}.
\end{thebibliography}

\bibliographystyle{unsrtnat}
\end{document}